\documentclass[conference]{IEEEtran}
\IEEEoverridecommandlockouts\usepackage[utf8]{inputenc}
\usepackage{amsmath}
\usepackage{amsfonts}
\usepackage{amssymb}
\usepackage{amsthm}
\usepackage{color}
\usepackage{graphicx}
\usepackage{caption}
\usepackage{subcaption}
\graphicspath{ {./newDrawings/} }
\usepackage{textcomp}
\usepackage{xcolor}
\def\BibTeX{{\rm B\kern-.05em{\sc i\kern-.025em b}\kern-.08em
    T\kern-.1667em\lower.7ex\hbox{E}\kern-.125emX}}

\usepackage{algorithm}
\usepackage[noend]{algpseudocode}

\algnewcommand\algorithmicswitch{\textbf{switch}}
\algnewcommand\algorithmiccase{\textbf{case}}
\algnewcommand\algorithmicassert{\texttt{assert}}
\algnewcommand\Assert[1]{\State \algorithmicassert(#1)}%
\algdef{SE}[SWITCH]{Switch}{EndSwitch}[1]{\algorithmicswitch\ #1\ \algorithmicdo}{\algorithmicend\ \algorithmicswitch}%
\algdef{SE}[CASE]{Case}{EndCase}[1]{\algorithmiccase\ #1}{\algorithmicend\ \algorithmiccase}%
\algtext*{EndSwitch}%
\algtext*{EndCase}%

\algdef{SE}[SUBALG]{Indent}{EndIndent}{}{\algorithmicend\ }%
\algtext*{Indent}
\algtext*{EndIndent}

\author{Gal}
\title{Redacting Transactions from Execute-Order-Validate Blockchains}

\newcommand{\txid}{\emph{TXid}}
\newcommand{\hlf}{Hyperledger Fabric}

\newtheorem{remark}{Claim}

\author{\IEEEauthorblockN{Yacov Manevich}
\IEEEauthorblockA{\textit{IBM Haifa Research Lab} \\
\textit{Haifa, Israel}\\
yacovm@il.ibm.com}
\and
\IEEEauthorblockN{Artem Barger}
\IEEEauthorblockA{\textit{IBM Haifa Research Lab} \\
\textit{Haifa, Israel}\\
bartem@il.ibm.com}
\and
\IEEEauthorblockN{Gal Assa \thanks{$\dagger$ This  work  was  done  when  Gal Assa was  an  intern with IBM Haifa Research Lab}}
\IEEEauthorblockA{\textit{Electrical Engineering} \\
\textit{Technion, Isreal Institute of Technology $^\dagger$}\\
Haifa, Israel \\
galassa@campus.technion.ac.il}
}
\begin{document}
\maketitle

\begin{abstract}
As user privacy gains popularity and attention, and starts to shape relations between users and service providers, blockchain based solutions thrive for ways to relax immutability without sacrificing consistency. This work answers that need and presents the first design for a redactable execute-order-validate blockchain, that grants users with the \emph{right to be forgotten}. The design is easy to adopt, as we exemplify by implementing it on top of Hyperledger Fabric. It modifies the block structure and extracts user data from the hash-chain without loosening any correctness or liveness criteria. We evaluate our design and show that it provides compliance with only a minimal performance overhead, making it a feasible add-on to any execute-order-validate blockchain system. 
\end{abstract}


\section{Introduction}
User privacy has been gaining increasing attention from service providers, general public, and lawmakers as more and more services migrate their data to the cloud, and users increasingly rely on service providers to safeguard their personal information. General Data Privacy Regulation (GDPR) and California Consumer Privacy Act (CCPA) are well known representatives out of many laws, acts, and regulations which define the relations between service users and providers, and deal with the very fundamental question: \emph{Who owns user data?} The EU via GDPR \cite{parliament2016regulation}, as well as California via CCPA and other states and counties believe that personal data belongs to users, and therefore require service providers to grant their users with the \textit{right to be forgotten}, i.e., user data should be deleted upon the user's request. This work suggests a practical way to satisfy the latter requirement in the context of permissioned blockchains.

\paragraph{The right to be forgotten}
As data ownership shifts towards users, one may choose to remove their information from some service provider's data store. From deleting (\textbf{actually deleting} -- not only making invisible to other users) a picture in a social network to removing all activity on a bank account upon closing the account, the general public is now allowed to ask service providers to permanently remove their data, and be granted. 

Organizations and service providers willing to comply with data privacy regulations ought to be able to locate, collect and remove the personal data of any specific user upon request. This introduces challenges to providers who store data. How is the data removal operation defined? How to make sure the integrity of the remaining data remains intact after data is removed? How to effectively trace back the derivation of a record when its history is susceptible to erasure? We try to provide answers to these questions in the context of a technology whose fundamental property is history being immutable, \emph{blockchain}.

\paragraph{Blockchain technologies are here to stay} Blockchains are essentially distributed, verifiable ledgers. In recent years, blockchains increased its popularity as a means of payment and asset transfer.  Public opinion towards blockchain has changed over the years, at first, the general public was either unaware or suspicious, later the public became skeptical, and nowadays blockchain is seen as the future of decentralized asset management. Quite surprisingly, this belief is shared between believers in the traditional banking and financial systems, and visionaries who advocate for replacing these exact systems with open, unmediated, peer-to-peer ones.  The understanding that there is a technology that allows agreement with neither mutual trust nor a central mediator drives governments, banks, and numerous companies to invest in private, public, or consortium blockchains. Alongside trustless agreement, many blockchains also provide visibility of the complete history of the system, via a shared and distributed ledger -- more precisely, an agreed upon prefix -- the system gains consistency, i.e., participants who join the system deduce its state according to the history they learn about. This means that participants observing the same history, see the system in the same state. By all participants and observers knowing \textit{who} did \textit{what} and \textit{when}, systems gain \textit{auditability},
 hence participants are held responsible for their set of actions throughout the entire history, as every action can be verified against the logic of the system, and every asset may be tracked back up to its creation.

\begin{figure}[ht]
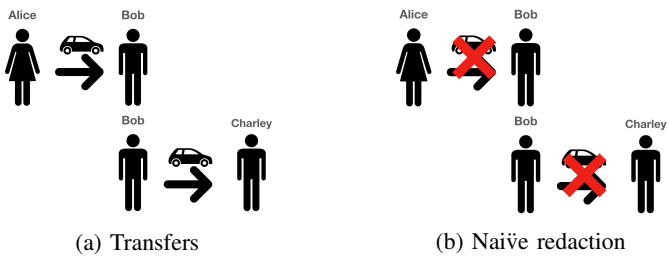

\centering
\begin{subfigure}{.2\textwidth}
	\includegraphics[width=\textwidth]{alice1.jpeg}
	\caption{Transfers}
	\label{fig:alice1}
\end{subfigure}
\hfill
\begin{subfigure}{.2\textwidth}
	\centering
	\includegraphics[width=\textwidth]{alice2.jpeg}
	\caption{Nai\"ve redaction}
	\label{fig:alice2}
\end{subfigure}
\caption{Redaction of asset transfer transactions}
\label{fig:aliceBob}
\end{figure}

\paragraph{Append-only structures never forget}
Originally, blockchains were designed as distributed ledgers, immutable logs of records, append-only structures by nature. The right to be forgotten seems to be in conflict with this nature, and makes redaction a non-trivial task. Designing a redaction mechanism should be done carefully: What should be deleted when a user asks to be deleted from the system? A na\"ive solution such as deleting the traces of the user's activity from the chain is likely to interfere with the auditability property, and lead to inconsistencies between existing nodes and newcomers. Let alone the fact assets may seem to appear and disappear on arbitrary occasions, joining nodes need to obtain the system's current state upon arrival based on the content of the distributed ledger. Replaying transactions in presence of redaction may lead to the following scenario, shown on Figure \ref{fig:aliceBob} Assume Alice transferred a car to Bob, who later transferred the same asset to Charley (Figure \ref{fig:alice1}). After these transactions were recorded on the ledger, Bob wants to be forgotten. If all the transactions Bob was involved in are redacted (Figure \ref{fig:alice2}), a new node will determine that the car is still possessed by Alice, whereas it is actually owned by Charley. This example emphasizes the difference between redaction of data and reverting transactions. 

\paragraph{Architecture}
We focus our solution on permissioned blockchains, whose core difference from their premissionless counterparts it the use of a public key infrastructure for authentication. We concentrate particularly in those of the \textit{execute-order-validate} architecture (pre-order execution). We find this setting most relevant since (1) permissioned blockchains are excel at modeling business agreements between companies, or capturing the business relation between end users in the presence of service providers and regulators; and (2) the \textit{execute-order-validate} paradigm improves on the predominant \textit{order-execute} (namely, post-order execution) architecture, by enabling concurrent processing of transaction and therefore increasing throughput and scalability. Breaking the throughput barrier is a critical step towards enabling wide use of blockchains in industrial applications, and many attempts \cite{eyal2016bitcoin,gilad2017algorand,kogias2016enhancing} have been made to design blockchain systems with throughput of hundreds and thousands of transactions per second. For reference, Visa's peak throughput requirement amounts to over 65,000 transactions per second \cite{Visa}. 


Our design is evaluated on Hyperledger Fabric (Fabric, for short) ~\cite{hyperledger}. It is an extensible permissioned blockchain platform which acts as a distributed operating system, and is implemented in the execute-order-validate architecture. Fabric provides high throughput and shows good scalability~\cite{8946222}, both are achieved through concurrent and optimistic transaction execution in a flexible trust model. Moreover, Fabric is able to cope with undeterministic smart contracts and transactions, and prevent their side effects. Smart contracts can be programmed in popular languages like Go and Java. These features make  Fabric a good candidate for wide adoption. Indeed, Fabric enjoys a wide community of developers with thousands of GitHub forks, and is currently used by tens of service providers with thousands of clients.

\paragraph{Motivation}
This work is motivated by the fact that certain privacy regulations are to be met in order to diversify the use of blockchains as cornerstones of services provided to consumers and businesses, as blockchains allow achieving  mutual goals without full trust. 

For instance, decentralized finance (\emph{De-Fi}) applications gain momentum and establish a role as a medium for peer to peer loans and investment opportunities with high interest rates. The increasing popularity of these services may affect systems that are driven by cryptocurrency and inflate transaction fees towards sums that are inaccessible to the common users. In fact, permissioned blockchains may be extended further to facilitate credit score for users and organizations, and providing such services in an authenticated system requires ability to redact user data from the system in order to comply with privacy regulations.




\paragraph{Paper organization}
The rest of the paper is organized as follows: Section \ref{background} provides the necessary background to this work. In Section \ref{design} we describe our design for general \textit{execute-order-validate} blockchains, and use Hyperledger Fabric as a concrete example for which we specify actual modifications to the system. We then evaluate the impact of our changes on Hyperledger Fabric in Section \ref{evaluation}. Section \ref{relatedWork} overviews related work. Section \ref{discussion} concludes the paper. 

\section{Background}\label{background}

In this section, a short introduction to blockchain is followed by the preliminaries regarding the cryptography in use. Further on we elaborate on paradigms to execute transactions and smart contracts in blockchain systems and dive deeper into the essentials of Fabric.



\subsection{Blockchain}
Blockchains are divided into two classes, permissioned and permissionless. The core difference between these classes is whether or not the membership of the system is controlled by some entity: permissionless blockchains, like Bitcoin \cite{nakamotobitcoin}, Ethereum \cite{buterin2014next}, and Zcash \cite{zcash} allow participants to join the system freely without specifying their identity, whereas permissioned blockchains identify the participants, and may treat them differently based on identity.

Permissionless blockchains often accommodate cryptocurrencies, whereas permissioned ones are used to settle and record agreements or track supply chains, by potentially competing organizations which do not fully trust one another. Permissioned blockchains have gained a fair bit of attention in recent years, and are investigated in the context of consensus protocols ~\cite{buchman2016tendermint, hotstuff} as well as full-fledged systems like Hyperledger Fabric ~\cite{hyperledger} and Tezos ~\cite{tezos}. As our interest is in the relation between users and service providers, this work addresses permissioned blockchains.

Blocks on the chain contain evidence for the execution of simple transactions or smart contracts. Smart contracts are code segments that are obligatory once agreed upon. They may be used to achieve long-term goals, bind transactions that ought to be executed in some order, or enforce the fulfillment of guarantees between participants. For example, if Alice wants to buy a car from Bob, but wants to pay only when the car is supplied, and bob wants to be sure that he has paid upon delivery a smart contract can enforce the rule that payment will be made if and only if the car was supplied.

\subsection{Cryptographic preliminaries}

The chain of blocks is linked by hash values. Every block points to its predecessor by the latter's hash. Hash values are calculated using \textit{hash functions}, which map arbitrary sized strings to fixed sized string.  Typically, the hash function in use are cryptographic, meaning they satisfy the following properties:

\begin{enumerate}
\item One wayness, which states that given a random output $\mathit{y}$ from the image of the function $h$ it is hard to find an input $\mathit{x}$ to $h$ such that $h(\mathit{x})=\mathit{y}$, and also that given an input $\mathit{x}$ it is hard to find an input $\mathit{x'} \neq \mathit{x}$ such that $h(\mathit{x}) = h(\mathit{x'})$.
\item Collision resistance, which extends one wayness with the additional requirement that it is hard to find two inputs $\mathit{x_1}$ and $\mathit{x_2}$ such that $h(\mathit{x_1}) = h(\mathit{x_2})$ 
\end{enumerate}
Informally, hard to find means the probability of finding such value or values within a given amount of time is negligible. These properties are essentially the main enablers of cryptocurrency systems, as they allow easy validation and at the same time keep forgery practically impossible as long a computationally efficient way to generate collisions remains unknown. There are, however, some functions, called chameleon hash functions \cite{Chameleon}, that enable efficient generation of collisions using a trapdoor. Without knowledge of the trapdoor, the aforementioned properties are preserved. 


\subsection{Order execute paradigm}
Presented in Fabric\cite{hyperledger}, execute-order-validate architecture consists of three phases. In the execution phase,  chaincode, i.e., transactions, smart contracts, or configuration code, is executed in isolation and its effects are recorded. The records consist of the information required to determine the validity of the execution's output before commit, and the actual values to be committed. To accommodate early, optimistic execution, a snapshot of the system's state is to be maintained at each peer. The state is usually maintained as a key-value store, called the \textit{Peer Transaction Manager (PTM)}. 



The output of the execution phase serves as input to the ordering phase. Ordering nodes run an atomic broadcast protocol between them, resulting in an ordered sequence of chaincode outputs grouped into blocks. Note that atomic broadcast and consensus are equivalent problems \cite{Chandra}. 

Broadcast of blocks with chaincode outputs reaches validator nodes, which validate it against the most recent system state. Once outputs are successfully validated they are committed and changes are reflected to the local states of the different peers. Not every result obtained via optimistic execution can be applied to the system, given the serialization order is dictated by the ordering service only after the output of the speculative execution is determined. For instance, consider the the next scenario with chaincode in the form of transactions. A committed concurrent transaction $\mathit{T}_1$ can cause another transaction $\mathit{T}_2$ to abort if it either brings the system to a state from which it is illegal to perform $\mathit{T}_2$ or if it writes a value to a key that $\mathit{T}_2$ has already read a previous version of.  
As transactions are validated and committed in the same order at all nodes, the consistency of the state machine replication process is guaranteed. 

Order-execute blockchains, on the other hand, determine execution order prior to execution, and in fact facilitate pessimistic concurrency control. The consensus takes place before the effect of transactions (or chaincode) is considered. The order-execute paradigm does not require additional data structures for maintaining a snapshot, but on the other hand does not benefit from storing data off-chain except for caching. This key difference plays an important role in our solution.

\subsection{A brief introduction to Hyperledger Fabric}
\begin{figure}[ht]
\begin{subfigure}{.5\textwidth}
 	\centering
 	\includegraphics[scale=.15]{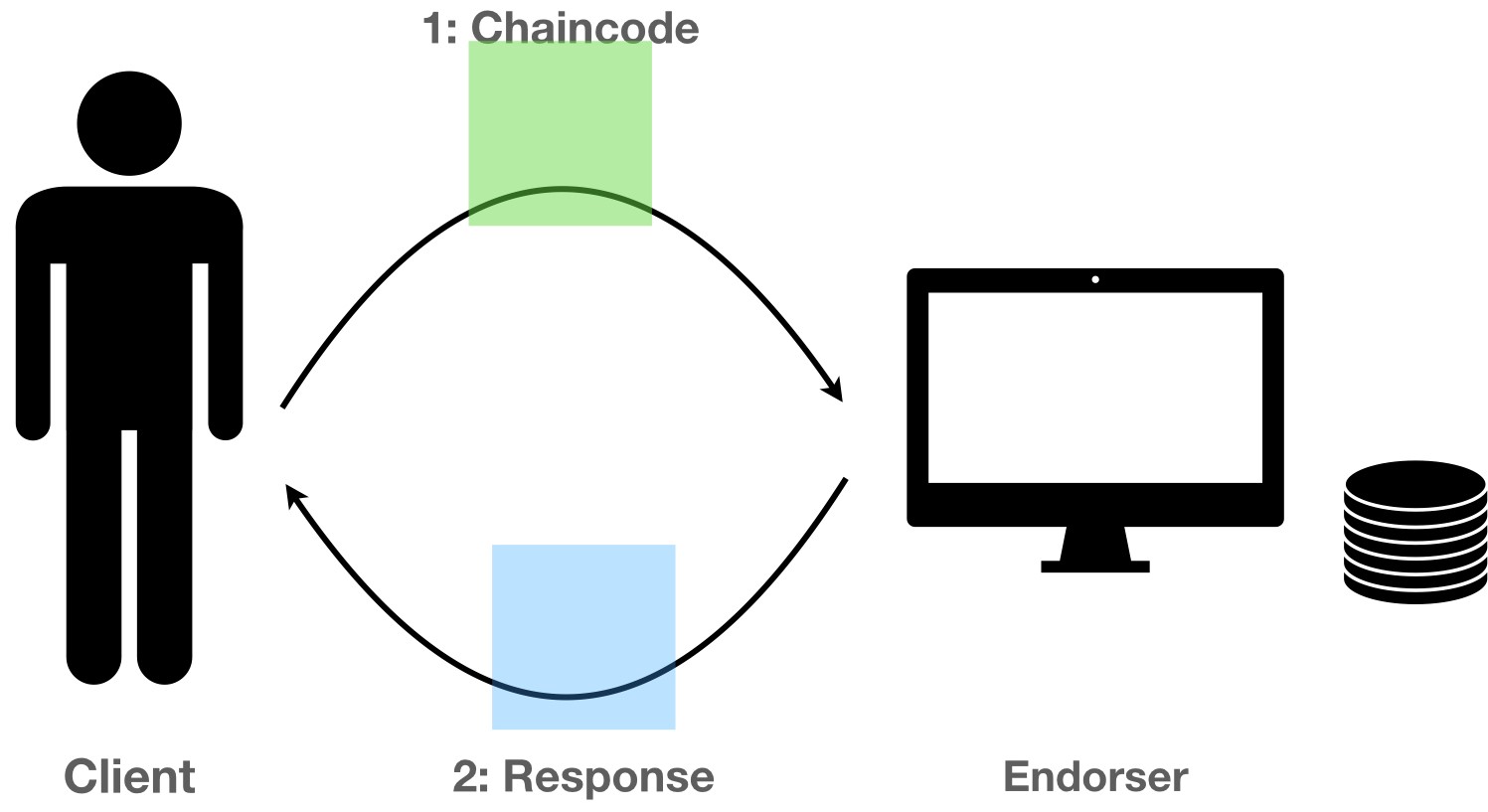}  
	\caption{Execution phase}
	\label{fig:execute}
\end{subfigure}
\begin{subfigure}{.5\textwidth}
	\centering
	\includegraphics[width=\textwidth]{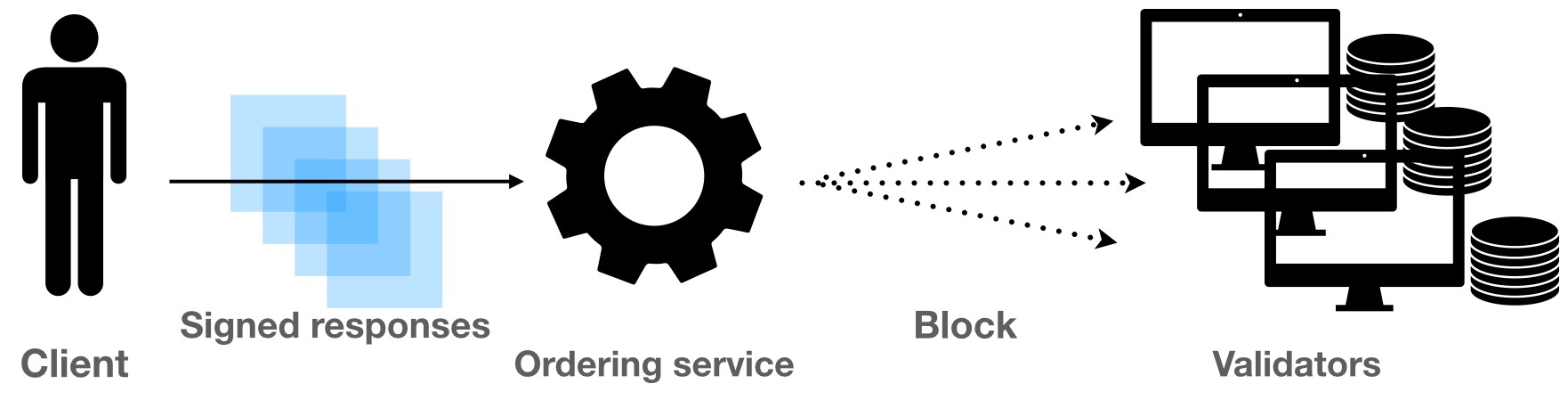}  
	\caption{Ordering and validation phases}
	\label{fig:orderValidate}
\end{subfigure}
\caption{Execute-order-validate scheme}
\label{fig:scheme}
\end{figure}

We now describe different components of the Hyperledger Fabric that have a role in the solution we propose. This work addresses clients, endorsers, orderers, and committers (namely, validators). Clients are participants in the system who initiate transactions. Transactions in Fabric execute fragments of chaincode atomically and in isolation (via Docker images). Transactions are sent to a set of Endorsers to be advocated by them. The flow is presented in figure \ref{fig:scheme}.

Endorsers optimistically execute (namely, simulate, without committing) transactions as requested by clients in an isolated manner, and sign a set of writes and a set of reads that define the result of executing a successful transaction, as seen on figure \ref{fig:execute}. These read- and write-sets are used later for validation and commit. The signed read- and write-sets  are sent back to the clients. When clients collect sufficient endorsements according to the endorsement policy, the output of the transaction, i.e., read- and write-sets, with the relevant signatures is sent to orderer nodes.

Figure \ref{fig:orderValidate} illustrates the flow that is initiated when a client collected enough endorsements and sends them to the ordering service. Orderer nodes order transactions and are the heart of the replication (namely, consensus) mechanism of the distributed ledger, as described previously. The ordering nodes form the ordering service, which totally orders transactions. Transactions are only ordered if their sets of signatures are valid and comply with the system's configuration and endorsement policy, which define who is required to endorse a given transaction. The orderers are stateless, and in particular ordering nodes do not access transaction contents. 

After transactions are ordered, their effect should be validated against the current state of the system. Although all endorsed transactions that comply with the system's policy are ordered, not all commit. Transactions failing to pass validation are discarded. This is the result of endorsers executing transactions optimistically, based on a previous snapshot of the system and before they are ordered. Conflicts like the one described above are detected by the validators. Transactions that successfully pass validation are committed and their corresponding key-value changes are applied to the database. Specifically, the changes are applied by the PTM, and all transactions are organized as blocks that appended to the ledger. Unlike orderers, validators read the keys in the write-sets of incoming transactions for concurrency control. Still, validators \emph{do not access} the values of the written keys as part of the validation process. This observation lead to key decisions in our design, as the endorsers' signatures suffice to guarantee correctness with regard to application logic.

Fabric guarantees snapshot isolation, meaning that a transaction commits only if it can appear to have operated on a snapshot of the system. Notice, however, that this does not require total order on all transactions. Therefore the use of multi-version concurrency control is fruitful, as transactions may commit even if they could not serialize by the total order determined by the ordering service.  In a nutshell, reads and writes performed by the transaction are collected to be used in the following manner: as part of the validation process reads are validated, in order to guarantee that upcoming writes could be performed without compromising transactional semantics, which require transactions to appear to have executed in a single instant. That means that if a value that was read by the transaction is changed before the transaction is validated, it is impossible to find a serialization point for the transaction and as a result the transactions aborts. 


\section{Design}\label{design}
In this section we describe the updates to be applied to a system built in execute-order-validate architecture. We describe changes to existing flows as well as the new redaction flow, and then exemplify via \hlf, to shed light on some implementation details. 

Execute-order-validate permissioned blockchains allow a rather straightforward approach towards redacting values from the ledger. For brevity, we assume a single key is written in each transaction. Extending it to support redaction of any subset of the keys written to in a transaction is straightforward. Our implementation on Fabric supports the extended capability. We assume that some unique identifier of the transaction (henceforth \txid) is known prior to redaction. 

Redacting transactions from a blockchain in a correct manner is not a trivial task, as shown in ~\cite{8728524,ateniese2017redactable,deuber2019redactable}. Modifying a block is, from a practical point of view, equivalent to changing its hash, hence in a way breaking the validity of the blockchain.
We do not want to tamper with the hashchain, and at the same time we seek for a solution that is thrifty in terms of both computation and storage, namely, a design that is robust and has low overhead. 
To achieve this, we rely on the collision resistance property of standard cryptographic hash functions in use. It allows the calculation of the block hash over the hashes of transactions content rather than over the actual content. Moving the redactable data out of the input to the block hash computation allows removing it without forgoing chain consistency. In the next subsection we describe the modifications to existing transactions, and the following subsection presents a realization on \hlf. 

\paragraph{Problem statement}
We are ready to formalize the redaction problem for permissioned blockchains. Our requirements are as follows: \begin{enumerate}
	\item Successful redaction results in the deletion of the requested data from the blockchain. Note that redaction does not revert the effect of the redacted transaction.
	\item Nodes joining the system will not be able to observe the contents of redacted data on the blockchain, had they received it from a correct node.
	\item Nodes joining late should be able to validate and ensure integrity of the hash chain and correctness of the data.
\end{enumerate}

\paragraph{Model and cryptographic assumptions}
The model is inherited from Hyperledger Fabric \cite{hyperledger}. Our trust model assumes any client may be Byzantine (i.e., can act maliciously or arbitrarily), and that peers are divided in groups: peers in the same group may trust each other, whereas peers from different groups do not. Groups may represent organizations or other abstractions. The design relies on an operating Byzantine (or crash) fault tolerant consensus algorithm. Our solution considers both crash-recovery and Byzantine failures of peers. 


\subsection{Redaction in execute-order-validate blockchains}

In execute-order-validate blockchains, transactions go through the process of being simulated (i.e., endorsed), next they are ordered in some total order, and then verified before finally being reflected to the world state as represented locally on each participating node. We recall that hashes are validated upon ordering and validation. Aiming at leaving the hashchain intact while allowing redaction, we now specify the following steps to be taken, and explain how auditability and consistency are preserved.

\begin{figure}[ht]

\begin{subfigure}{.45\columnwidth}
\centering
 	\includegraphics[width=.6\columnwidth]{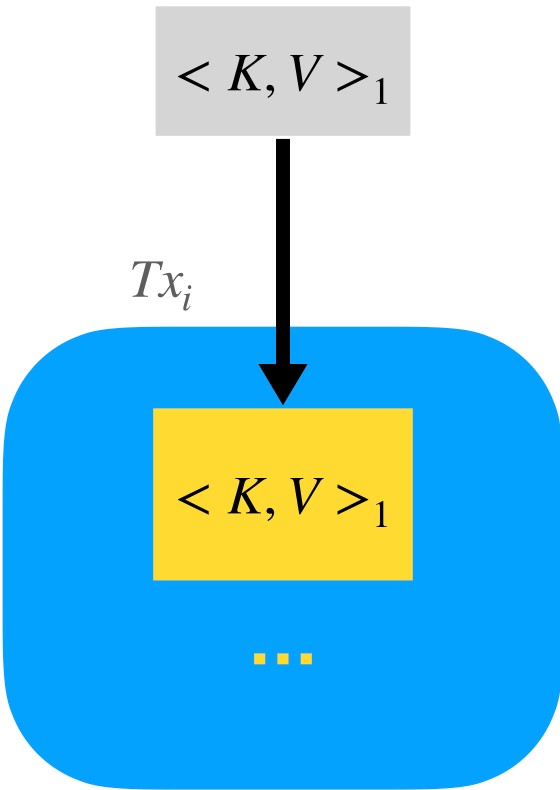}  
	\caption{Original transaction structure}
	\label{fig:b4}
\end{subfigure}
\hfill
\begin{subfigure}{.45\columnwidth}
	\centering
	\includegraphics[width=\columnwidth]{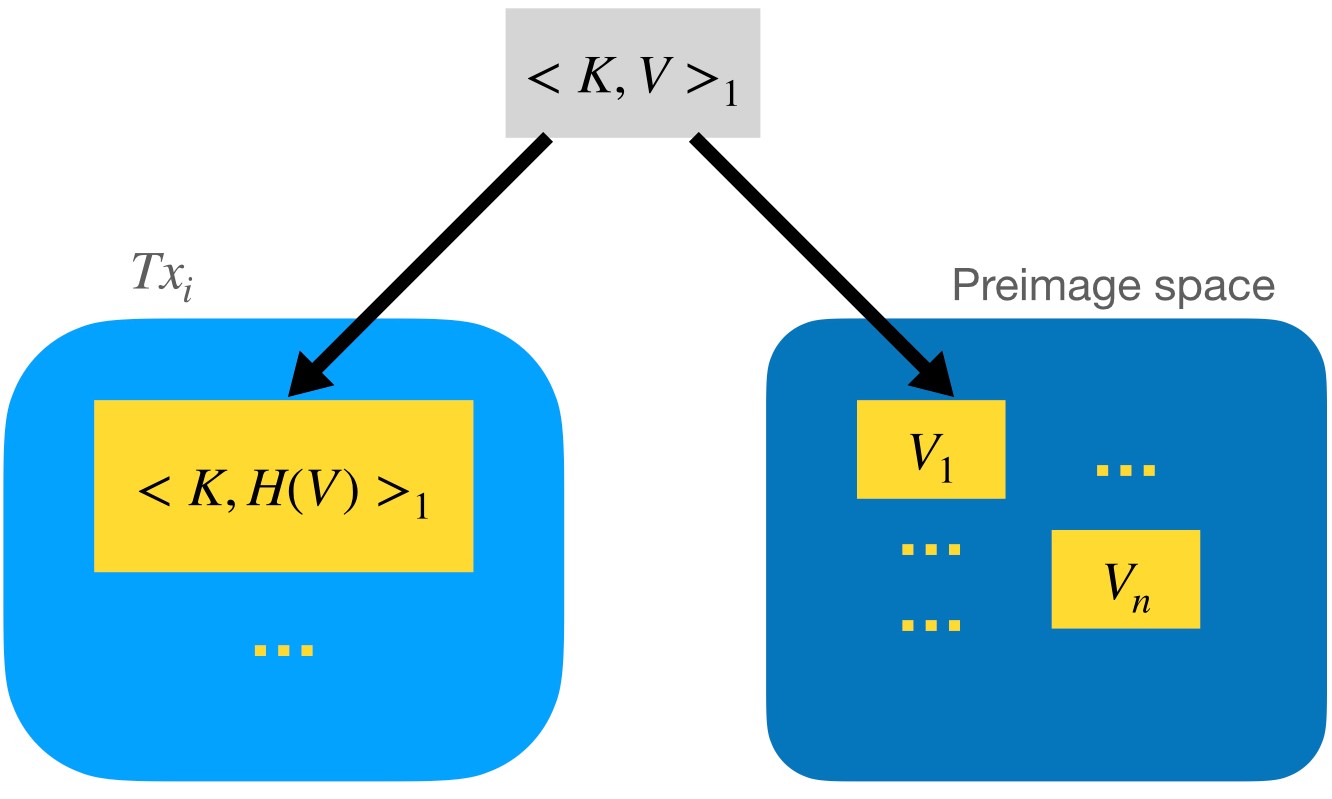}  
	\caption{New transaction structure}
	\label{fig:after}
\end{subfigure}
\caption{Write sets before and after change}
\label{fig:txStructure}
\end{figure}

\paragraph{Decouple user-data from hash-chain} 
The Blockchain needs to remain verifiable. Our design goal is to maintain this property without either recalculating hash-chains or keeping multiple versions of block contents as both of these overheads can be avoided. Figures \ref{fig:b4} and \ref{fig:after} illustrate the transaction structure before and after the change, respectively. Notice that in the prior form, transaction information is a first-class member of the block and is hashed as part of the hashchain. Our modification does not move sensitive data out of the block. Rather, the sensitive data resides in the block, but in a dedicated space which is not taken into account when hashing the block or signing it. The change makes redaction simple from the validation perspective, as we show next. We note that it is the responsibility of the user to ensure that its confidential data has sufficient entropy, as the one-wayness property of the hash function only holds for input sampled from a large distribution. This is often achieved by addition  of "salt" which is random bytes and is done in the application layer of te smart contract, making it practically impossible to find a preimage given the hash.

\paragraph{Place user data on a dedicated segment}
Hashes alone do not suffice when reconstructing the global state, thus actual transaction data ought to be stored and available. In order to bind the transaction data to the block, we collect it in a dedicated space, and replace the actual data in the segments that previously had user data with their hashes. The dedicated space for pre-images is shared among all transactions in the block and is illustrated by the rightmost container in figure \ref{fig:after}. 

\begin{figure}[htb]
\centering
\includegraphics[scale=.2]{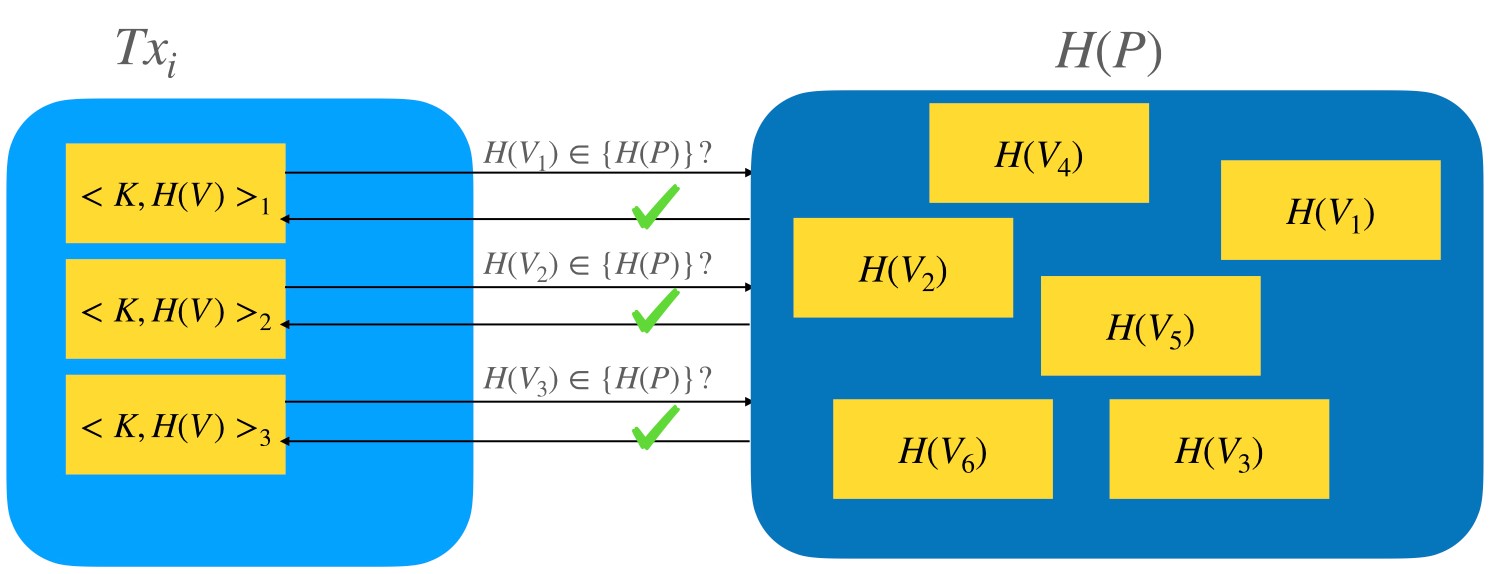}
\caption{Block validation logic}
\label{fig:validate_nominal}
\end{figure}

\paragraph{Validation} 
Validation now has an additional step which is described in figure \ref{fig:validate_nominal}. Observe that modified values within every transactions are hashed and matched against the set of hashes over all preimages in the block, denoted $H(P)$. In case a value's hash is not matched, the block is declared invalid. Algorithm \ref{alg:val} shows the pseudo-code for block validation. The additional validation logic interacts with the preimages and their hashes. First, the preimages are hashed to a new set of hashes (line \ref{inalg:gather}), or counted redacted ($preimageRedactionCounter$) if they were marked as such. Note hashes in the set are not linked to specific transactions. Next, the hashes in each transaction's space are searched in the new set. In case they are not found, they are counted redacted ($hashMismatchCounter$). If the two counters match, the block is considered valid. 

Note that comparing counters is sufficient due to creating artificial mismatches being considered too difficult in terms of computation, so it is practically impossible to trick the algorithm with compensating hashes/preimages. It is also important to note that the comparison in the other direction (i.e., verifying every hash has a matching preimage) is unnecessary, as it is impossible to forge the content of a transaction due to the signatures by endorsers.

\begin{algorithm}
\caption{Addition to validate method}
\label{alg:val}
\begin{algorithmic}[1]\footnotesize

\Procedure{Validate}{block}
\State $\mathcal{H}$ $\gets \{\}$
\State preimageRedactionCounter $\gets$ 0
\State hashMismatchCounter $\gets$ 0
\ForAll {PreImage p $\in$ block.preImageSpace} \label{inalg:gather}
	\If {p == "0000...0"}
		\State preimageRedactionCounter++  \Comment{Redacted}
	\Else
		\State $\mathcal{H}$ $\gets$ $\mathcal{H}$ $\cup$ {$Hash(p)$}			
	\EndIf
\EndFor

\ForAll {Transaction $\tau$ $\in$ block}
	\If {$\tau$.hashes $\not\subseteq $ $\tau$.hashes $\cap$ HashedPreImages}
		\State $\Delta = $ $\tau$.hashes $\setminus$ $\tau$.hashes $\cap$ HashedPreImages
		\State hashMismatchCounter $\gets$ hashMismatchCounter + $|\Delta|$ 
	\EndIf 
\EndFor

\If{preimageRedactionCounter $\neq$ hashMismatchCounter}
	\State \Return ValidationError
\EndIf

\State \Return Success
\EndProcedure
\end{algorithmic}
\end{algorithm}

\subsection{Redacting transactions}

We are finally ready to present the redaction mechanism. Redaction is done using a dedicated instruction which results, upon its commit, in deletion of the data of the redacted transaction from peers' ledgers. Instead of deleting the data completely, we replace it with a zero-value. This helps distinguishing missing values from redacted ones, and retains the block size in file storage.


Recall that blocks contain hashes instead of transaction data now. The aforementioned action creates a mismatch between the actual content and its alleged hash. 
An observer, or a new peer, traverses the blockchain and validates it to generate a its own consistent world view. Since replaying regular and redaction transactions results in the observer observing redacted writes, the proposed mechanism overwrites values in a manner that retains validation, thanks to the fact that the value of a key write is not used in transaction validation. 

Deleting all data that involves some specific user from the blockchain is simple, and only requires locating all transactions writing to keys associated with that user and redacting them one by one using the specified mechanism. The associated transactions can easily be found using an index, a common component for data stores in general and key-value stores in particular. 

\paragraph{Preserving consistency in spite of data redaction}
A new node who joins the system and acquires the ledger, is unable to apply state updates that have been redacted, but it still has knowledge of the key that was redacted (just not its value). Therefore, to retain correctness in smart contract execution, keys that their corresponding values have been redacted, are considered as \emph{crippled keys} and whenever a smart contract queries such keys, the execution immediately aborts. We note that this is a temporary state, as we expect these keys to be either deleted or overwritten in a later block.


\paragraph{Validating redaction instruction}
We introduce a new type of transactions into the system which contain redaction instructions. Unlike existing transaction types, whose changes span the current state and the block they are included in, redaction transactions change a block which is already a part of the system's history, and does not change the value of any key as a side effect. Therefore, redaction instructions requires its own custom validation logic. 

First, redaction instructions should be signed and checked for sufficient signatures like other transactions. After being ordered, there is no need to check the content's validity beyond the existence of the transaction to be redacted on the blockchain. Once it is found, committing simply means removing the preimages from the dedicated preimage space of the relevant block. 

\paragraph{Validating a blockchain in presence of redacted transactions}
The validation procedure performed by observers and new joining peers is the same as existing peers, as described in Algorithm \ref{alg:val}. This is a design goal, since it is possible for existing peers to also to encounter a transaction for the first time after is had already been redacted. As transactions are validated against the preimage space in their validation phase, they go through a similar process upon validating an entire blockchain by observers. However, an observer may find hashes without corresponding preimages. This indicates redacted transactions. We note that it is not possible to forge a blocks content, as blocks are signed by orderers. It also leaves the validation task simple to perform, as a block will only be considered invalid in the rare case when the number of redacted preimages does not match the number of unmatched hashes in a signed block. Upon detecting a transaction whose hashes miss corresponding preimages, the observer learns that the transaction was redacted. Moreover, it is impossible for them to tell the content of a redacted transaction. 

\begin{figure}[ht]
\begin{subfigure}{.5\textwidth}
 	\includegraphics[width=\textwidth]{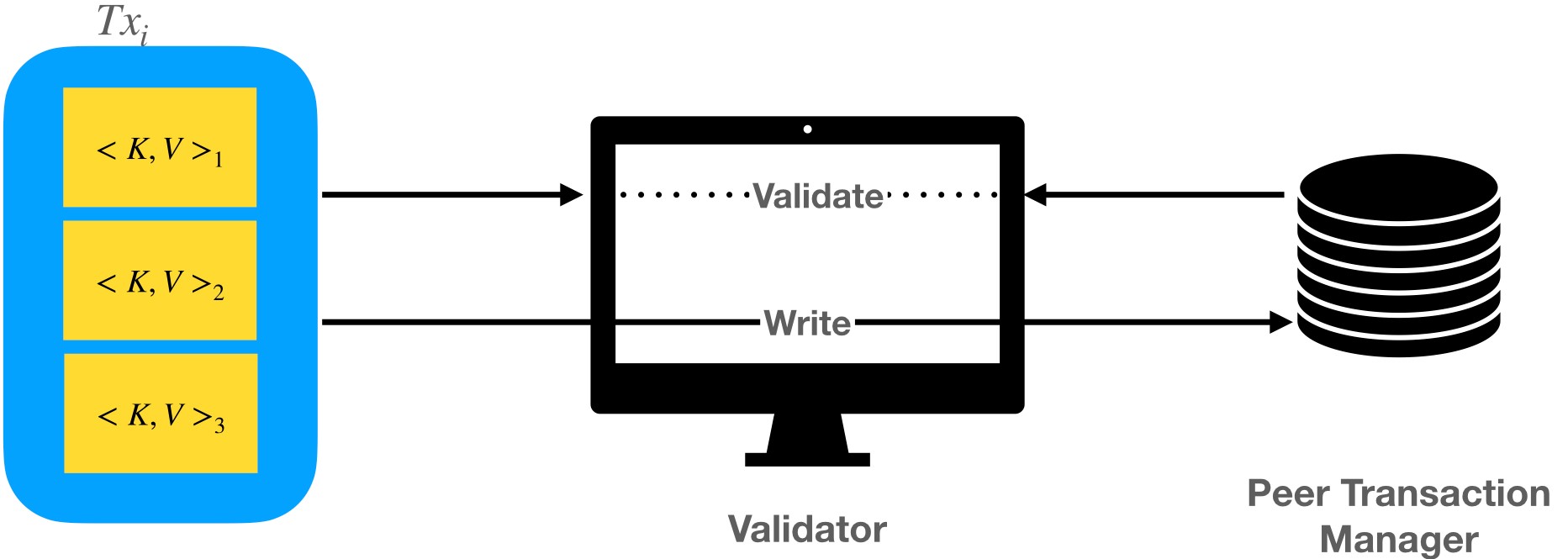}  
	\caption{Validation for append-only systems}
	\label{fig:val_before}
\end{subfigure}
\begin{subfigure}{.5\textwidth}
	\includegraphics[width=\textwidth]{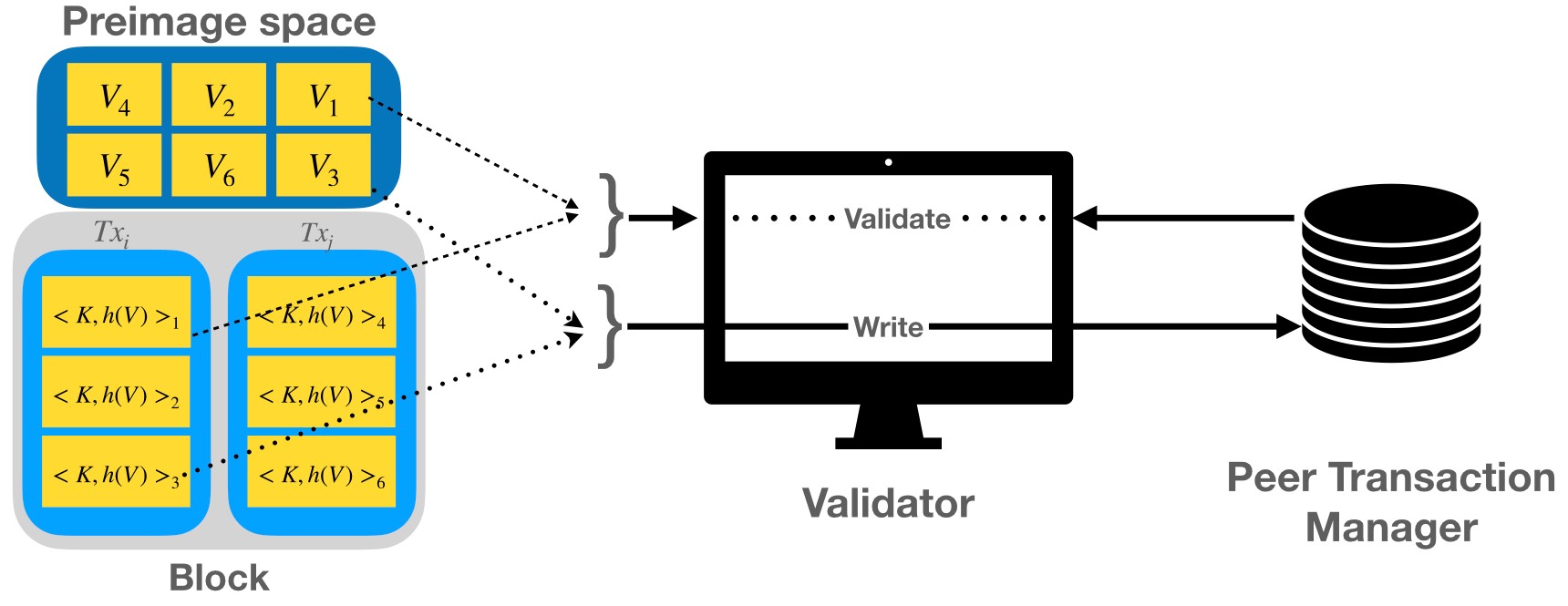}  
	\caption{Redaction supporting validation}
	\label{fig:val_after}
\end{subfigure}
\caption{Transaction validation}
\label{fig:validation}
\end{figure}

Figure \ref{fig:validation} shows the changes to validating nodes. In append only execute-order-validate blockchains, key-value pairs are present on blocks in their plain form. Matching keys to values is not required, so both validation and commit can be performed directly, as figure \ref{fig:val_before} shows. Supporting redaction requires the use of the preimage space, which in turn requires reassembly of key-value pairs prior to validation and commit, as figure \ref{fig:val_after} describes.

\paragraph{Consistency}

Peers maintain local structures, such as key-value stores, in order to capture snapshot of the entire system state up to the certain position of the ledger, the most recent update to a key's value suffices to determine its correct state. In particular, a joining peer who observes a subset of the system's history will be able to determine the correctness of the state they observe. We now formalize and prove the next claim:

\begin{remark}
Redaction never causes divergence in the ledger. A successfully validated ledger represents a consistent state. 
\end{remark}

\begin{proof}
We address consistency in two planess: blockchain, and state (PTM). First, note the integrity of the hash chain is guaranteed, as redaction does not have side effects on the hashed content, as the block hash computation disregards the pre-images.  On the PTM level, values are retrieved via corresponding writes on the blockchain. If the write was redacted, its value may be inaccessible to a new peer joining the network. However, as the key is declared crippled (III-B-a), executions that attempt to read it result in instant abort, satisfying the consistency criteria vacuously for such cases.

It is now left to show that redaction does not create contradicting views. To prove that, we assume by contradiction that two peers reached the same ledger height, however one peer joined before a redaction, and the other after the redaction took place have validated the same ledger but reached different states. In that case, there is at least one transaction the peers do not agree on, and in particular there is one that is the first, as transactions are ordered. That transaction cannot be a redaction transaction, because the ordering service only puts into blocks valid redaction transactions. Hence, the transaction must be an endorser transaction that has been redacted. From Algorithm 1, it follows that if a block has been declared valid as a whole, then all hash mismatches correspond to redacted transactions. Hence, from the construction of Fabric validaiton logic, both peers deem the transaction as valid or invalid discounting the redaction which removed the pre-images. Otherwise, Algorithm 1 outputted a validation error for the entire block, and that block is never committed in the peer that joined after the redaction, in which case its ledger height is not equal to the ledger height of the other peer, in contradiction to the assumption. 
\end{proof}


\subsection{Redactable \hlf}
We now detail the adjustments and modifications to \hlf~ in order to enable redaction in the system. We introduce the structure of redaction transactions and describe the required changes to Fabric's components in the nominal flow and the redaction flow. Note that our design leaves only a small footprint on Fabric's codebase.

\subsubsection{Nominal flow}
The client is oblivious to the design change, except the different structure of responses from endorsers. While the client does not experience changes, application chaincode does change a bit, and a salt is generated per written key at the client and sent to the endorser. The salt serves the purpose of making redacted data unrecoverable, and does not end up in the transaction sent to the orderer. Endorsers, however, have additional tasks. Aside from executing transactions, endorsers need to replace what was previously the values with their corresponding hashes, and return the pre-images additionally to the rest of the signed output (endorsement). After receiving sufficient endorsements, the client sends the transaction containing the endorsements to the ordering service.

Ordering nodes form blocks after validating the hashes and preimages match for each endorsed transactions, which is an addition to their original role. Blocks are formed differently, as hashes are calculated over different data, and the dedicated preimage space is attached to each block. The ordering task itself does not change, as it is agnostic to transaction contents. Finally, blocks are propagated to committing peers who in addition to their existing tasks, verify that the every hash of a value in the transaction's write-set has a matching preimage in the dedicated space. Committing transactions does not change significantly, except the source of the values to be written in the PTM and the structure of the blocks to be appended to the ledger.

\subsubsection{Redaction flow}
A client generates a redaction transaction, with the relevant \emph{TXid} and key to be redacted. This only happens after the value was actually deleted from the database representing the world state. The ordering service validates the redaction transaction (request) against a predefined policy, namely that enough parties request for the redaction, and if so, includes it in some future block. Finally, upon validation, both orderers and peers zero the preimages corresponding to \emph{TXid} in the relevant block in their ledgers. We assume redaction from the blockchain happens after the bespoken keys are either removed or overwrritten, and say that if it is not the case, then consistency in transaction validation is preserved anyway across all peers, however peers that join the network after the redaction will never acquire the values until the users overwrite the keys at a later time.

An observer or a peer joining the system may validate the blockchain exactly as described above, and distinguish malformed or tampered with blocks from correct ones with redacted writes using the same logic: A written value is considered redacted if no preimage in the dedicated space is found to match the hash associated with its key and there exists evidence for redaction in the block. Note that the hash-chain remains unchanged even in presence of redacted transactions, as the contents of the block which are input to the hash function are unchanged.

%
%
%

\section{Evaluation}\label{evaluation}

We evaluate our design via microbenchmarks on top of Hyperledger Fabric v2.2 on a 16 core Intel Xeon E5-2683 2.00GHz with 32GB RAM and 300MB/s storage. The intent of using microbenchmarks is focusing on the bottleneck of the system, which is the peer commit path \cite{8526892}, rather then the transaction execution or the consensus. We microbenchmark the commit throughput of a peer against both supported database variants: GoLevelDB \cite{goLevelDB} and CouchDB \cite{couchDB}. We compare the throughput achieved by a Fabric peer before and after applying the changes from our design. Finally, we discuss the results and their implications.

We create a controlled environment: to eliminate irrelevant factors, we use generated blocks in advance and store them on the local file system. We generate blocks with varying sizes, totaling 1,000,000 transactions per trial. Each transaction contains five key writes, each to a random key out of a total key space of 10 keys each 16 bytes in size and a value of 32 bytes. The lack of read dependency eliminates multiversion concurrency control errors and thus makes every key write be applied to the database. 

\begin{figure}[h]
	\centering
	\includegraphics[width=.5\textwidth]{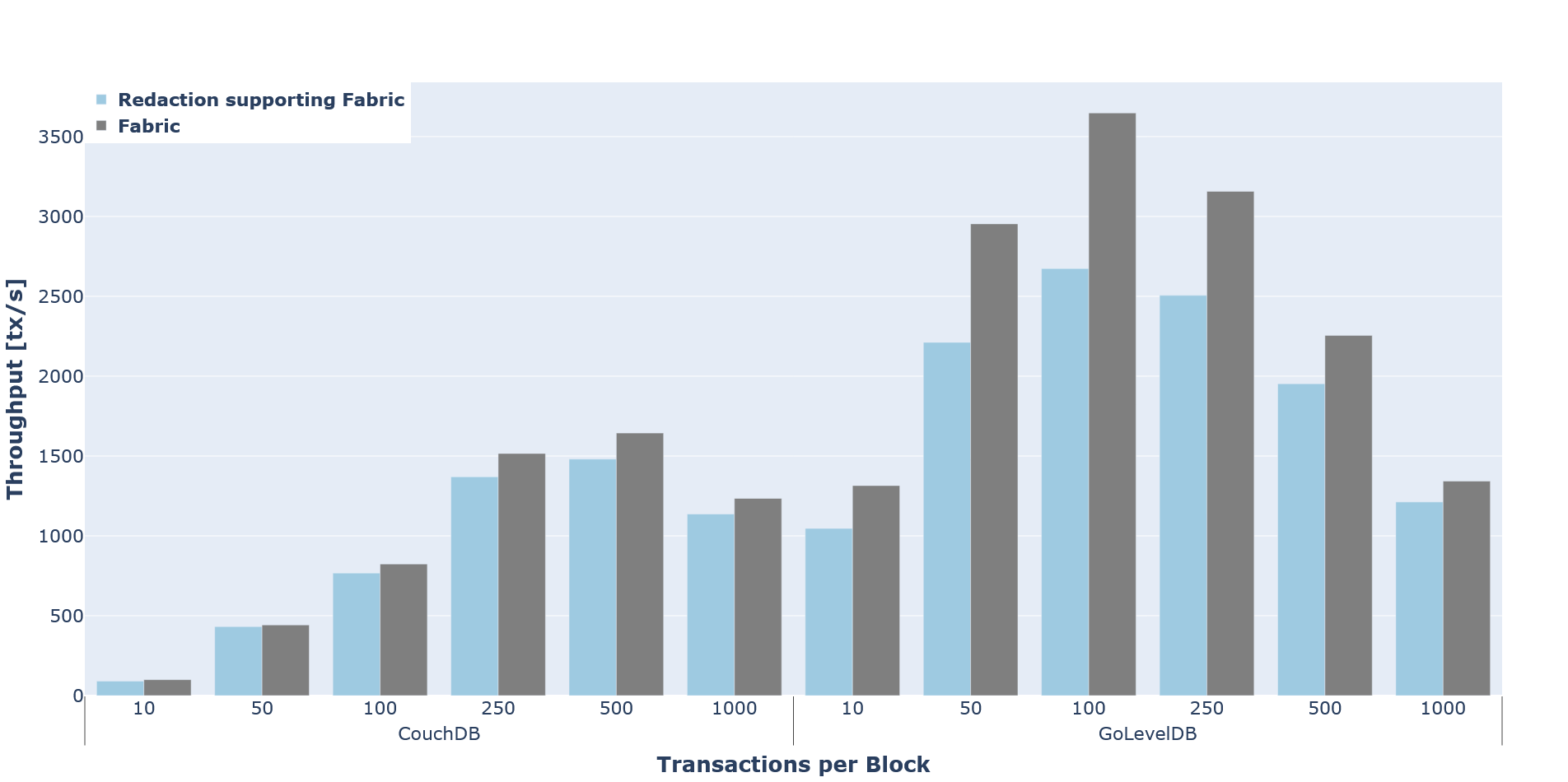}  
	\caption{Peer's Throughput}
	\label{fig:res_peer}
\end{figure}

Figure \ref{fig:res_peer} shows the throughput results before and after our design is applied, for varying number of transactions per block. Every data-point represents the average of 10 independent experiments. The performance overhead of applying our technique is 19\% for GoLevelDB, and 7\%  for CouchDB. We believe the performance penalty can be further reduced via optimizing the commit path, namely eliminating redundant data serializing and deserializing, similar to what was pointed out by \cite{FastFabric} as a major performance inhibitor. From a high level perspective, the commit of a block is a three stage process: (1) the block is parsed and each trasaction is validated; (2) the block is committed to the file system; and (3) the valid transactions are applied to the state database (GoLevelDB or CouchDB). The first and third stages both pre-process the block's deeply nested structure, and there is no sharing of intermediate results between the phases. To make our implementation as least intrusive as possible, we had to apply our techniques twice (once in each stage). We believe that refactoring the commit path that will eliminate redundant transaction structure processing, will not only improve the baseline performance, but also further diminish our processing overhead.




\section{Related Work}\label{relatedWork}

Multiple designs for mutable order-execute blockchains have been suggested in recent years. They vary in their techniques to modifying blocks, which in turn enable redacting transactions or modifying history.

We separate the various techniques by their high level approach: (a) Off-chain storage (b) Hash collision based; (c) Agreement based; (d) Ledger pruning driven.

\paragraph{Off chain storage approach} Troung et al. \cite{8876647} and Wirth et al. \cite{wirth2018privacy} suggest solutions where data is stored off-chain and the blockchain only manages pointers and hashes to the data. While these efforts may be GDPR compliant, they are orthogonal to this work, since the data itself never reaches the blockchain. 

\paragraph{Hash collision approach} One of the challenges in redacting information in a blockchain is that each block contains the hash of the previous block, thus a mutation in the previous block also changes the next block, which transitively changes all successive blocks. Chameleon hash functions, introduced by Krawczyk and Rabin \cite{Chameleon} are hash functions with a trapdoor that allows to compute a pre-image that results in hash collision. Employing a chameleon hash function trivally allows changing the transaction data while retaining the hash value of the next block.

However, any party that has knowledge of the trapdoor gains an excessive privilege since it can single-handedly rewrite history.
To that end,  Ateniese et al. \cite{ateniese2017redactable} devised a protocol which uses secure multi-party computation to keep the trapdoor shared across multiple parties, and computing the hash collision without any of the parties learning the trapdoor.  Ashrita et al. \cite{8728524} follow the same path of Ateniese et al. \cite{ateniese2017redactable}, and improve by using a non-linear secret sharing scheme and making trapdoor keys unique for each block, further amplifying the security. Using secure multi-party computation incurs a performance overhead that intensifies as the number of parties grows, and raises complexity concerns as new parties may join and leave the network. In contrast, our scheme does not require secure multi-party computation to accomodate redaction transactions, as these are processed seamlessly as regular ones.

\paragraph{Agreement approach} Deuber et al.  \cite{deuber2019redactable} handle redacting in the permissionless setting with voting in a dedicated consensus instance. In a way, our work is similar in the sense that redactions are agreed upon atomically broadcast (which is equivalent to consensus). However, their redaction mechanism cannot remove data that can be spent, such as transfer amounts. We note that this restriction is inherent to the order-execute nature of the blockchains they focus on.

\paragraph{Pruning approach} In addition to relaxing the immutability property of blockchains, some solutions have been proposed in order to compress the blockchain history into a agreed upon snapshot, after which the blockchain files can be erased, or kept for external auditability. The motivation is to ease joining new nodes into a network and reduce storage requirements. Examples can be found both in industry (Hyperledger Fabric's checkpoint proposal)\cite{Pruning} and academic works such as done by Palm et al. ~\cite{8525390}.

\section{Conclution} \label{discussion}
User privacy raises many questions and challenges in systems design. The tension between the right to be forgotten and the immutable nature of blockchains requires design flexibility in order to allow existing blockchain-based services to comply with privacy regulations and new services and applications to grow. Keeping performance, availability (which reflects as guarantees to end users, and in turn compliance with regulations), and ease of implementation in mind, we believe our design balances between the needs of service providers, application developers, and most importantly end-users.

Our results are indeed encouraging. We show that user privacy does not come at the price of system performance, when it comes to execute-order-validate blockchain systems, it is possible to comply with GDPR and similar regulations without the use of complex and advanced cryptographic techniques as in the state of the art \cite{8728524,ateniese2017redactable,deuber2019redactable}. We hope our design will remove privacy barriers that currently prevent organizations and service providers enjoy the benefits of publicly available blockchain solutions like Hyperledger Fabric. Our implementation is open source \footnote{https://github.com/FabricGDPR/} and is planned to be referenced in a future official proposal to the Fabric open source community.

\clearpage
\bibliographystyle{abbrv}
\bibliography{references}

\begin{thebibliography}{10}

\bibitem{hyperledger}
E.~Androulaki, A.~Barger, V.~Bortnikov, C.~Cachin, K.~Christidis, A.~De~Caro,
  D.~Enyeart, C.~Ferris, G.~Laventman, Y.~Manevich, et~al.
\newblock Hyperledger fabric: a distributed operating system for permissioned
  blockchains.
\newblock In {\em Proceedings of the thirteenth EuroSys conference}, pages
  1--15, 2018.

\bibitem{8728524}
K.~{Ashritha}, M.~{Sindhu}, and K.~V. {Lakshmy}.
\newblock Redactable blockchain using enhanced chameleon hash function.
\newblock In {\em 2019 5th International Conference on Advanced Computing
  Communication Systems (ICACCS)}, pages 323--328, 2019.

\bibitem{ateniese2017redactable}
G.~Ateniese, B.~Magri, D.~Venturi, and E.~Andrade.
\newblock Redactable blockchain--or--rewriting history in bitcoin and friends.
\newblock In {\em 2017 IEEE European Symposium on Security and Privacy
  (EuroS\&P)}, pages 111--126. IEEE, 2017.

\bibitem{buchman2016tendermint}
E.~Buchman.
\newblock {\em Tendermint: Byzantine fault tolerance in the age of
  blockchains}.
\newblock PhD thesis, 2016.

\bibitem{buterin2014next}
V.~Buterin et~al.
\newblock A next-generation smart contract and decentralized application
  platform.
\newblock {\em white paper}, 3(37), 2014.

\bibitem{Chandra}
T.~D. Chandra and S.~Toueg.
\newblock Unreliable failure detectors for reliable distributed systems.
\newblock {\em J. ACM}, 43(2):225–267, Mar. 1996.

\bibitem{deuber2019redactable}
D.~Deuber, B.~Magri, and S.~A.~K. Thyagarajan.
\newblock Redactable blockchain in the permissionless setting.
\newblock In {\em 2019 IEEE Symposium on Security and Privacy (SP)}, pages
  124--138. IEEE, 2019.

\bibitem{eyal2016bitcoin}
I.~Eyal, A.~E. Gencer, E.~G. Sirer, and R.~Van~Renesse.
\newblock Bitcoin-ng: A scalable blockchain protocol.
\newblock In {\em 13th $\{$USENIX$\}$ symposium on networked systems design and
  implementation ($\{$NSDI$\}$ 16)}, pages 45--59, 2016.

\bibitem{gilad2017algorand}
Y.~Gilad, R.~Hemo, S.~Micali, G.~Vlachos, and N.~Zeldovich.
\newblock Algorand: Scaling byzantine agreements for cryptocurrencies.
\newblock In {\em Proceedings of the 26th Symposium on Operating Systems
  Principles}, pages 51--68, 2017.

\bibitem{tezos}
L.~Goodman.
\newblock Tezos: A self-amending crypto-ledger position paper.
\newblock {\em Aug}, 3:2014, 2014.

\bibitem{zcash}
D.~Hopwood, S.~Bowe, T.~Hornby, and N.~Wilcox.
\newblock Zcash protocol specification.
\newblock {\em GitHub: San Francisco, CA, USA}, 2016.

\bibitem{kogias2016enhancing}
E.~K. Kogias, P.~Jovanovic, N.~Gailly, I.~Khoffi, L.~Gasser, and B.~Ford.
\newblock Enhancing bitcoin security and performance with strong consistency
  via collective signing.
\newblock In {\em 25th $\{$usenix$\}$ security symposium ($\{$usenix$\}$
  security 16)}, pages 279--296, 2016.

\bibitem{kraska2019schengendb}
T.~Kraska, M.~Stonebraker, M.~Brodie, S.~Servan-Schreiber, and D.~Weitzner.
\newblock Schengendb: A data protection database proposal.
\newblock In {\em Heterogeneous Data Management, Polystores, and Analytics for
  Healthcare}, pages 24--38. Springer, 2019.

\bibitem{krawczyk1998chameleon}
H.~Krawczyk and T.~Rabin.
\newblock Chameleon hashing and signatures.
\newblock 1998.

\bibitem{nakamotobitcoin}
S.~Nakamoto.
\newblock Bitcoin: A peer-to-peer electronic cash system.

\bibitem{8525390}
E.~{Palm}, O.~{Schelén}, and U.~{Bodin}.
\newblock Selective blockchain transaction pruning and state derivability.
\newblock In {\em 2018 Crypto Valley Conference on Blockchain Technology
  (CVCBT)}, pages 31--40, 2018.

\bibitem{parliament2016regulation}
E.~Parliament.
\newblock Regulation (eu) 2016/679 of the european parliament and of the coucil
  of 27 april 2016 on the protection of natural persons with regard to the
  processing of personal data and on the free movement of such data, and
  repealing directive 95/46/ec (general data protection regulation).
\newblock {\em Official Journal of the European Union L}, 119:1--88, 2016.

\bibitem{shastri13understanding}
S.~Shastri, V.~Banakar, M.~Wasserman, A.~Kumar, and V.~Chidambaram.
\newblock Understanding and benchmarking the impact of gdpr on database
  systems.
\newblock {\em Proceedings of the VLDB Endowment}, 13(7).

\bibitem{8876647}
N.~B. {Truong}, K.~{Sun}, G.~M. {Lee}, and Y.~{Guo}.
\newblock Gdpr-compliant personal data management: A blockchain-based solution.
\newblock {\em IEEE Transactions on Information Forensics and Security},
  15:1746--1761, 2020.

\bibitem{Visa}
Visa.
\newblock Visa fact sheet, 2018.

\bibitem{wirth2018privacy}
C.~Wirth and M.~Kolain.
\newblock Privacy by blockchain design: a blockchain-enabled gdpr-compliant
  approach for handling personal data.
\newblock In {\em Proceedings of 1st ERCIM Blockchain Workshop 2018}. European
  Society for Socially Embedded Technologies (EUSSET), 2018.

\bibitem{hotstuff}
M.~Yin, D.~Malkhi, M.~K. Reiter, G.~G. Gueta, and I.~Abraham.
\newblock Hotstuff: Bft consensus with linearity and responsiveness.
\newblock In {\em Proceedings of the 2019 ACM Symposium on Principles of
  Distributed Computing}, PODC '19, page 347–356, New York, NY, USA, 2019.
  Association for Computing Machinery.

\end{thebibliography}


\begin{thebibliography}{10}

\bibitem{Pruning}
Hyperledger fabric ledger checkpointing rfc
  https://hyperledger.github.io/fabric-rfcs/text/0000-ledger-checkpointing.html.

\bibitem{hyperledger}
E.~Androulaki, A.~Barger, V.~Bortnikov, C.~Cachin, K.~Christidis, A.~De~Caro,
  D.~Enyeart, C.~Ferris, G.~Laventman, Y.~Manevich, et~al.
\newblock Hyperledger fabric: a distributed operating system for permissioned
  blockchains.
\newblock In {\em Proceedings of the thirteenth EuroSys conference}, pages
  1--15, 2018.

\bibitem{8728524}
K.~{Ashritha}, M.~{Sindhu}, and K.~V. {Lakshmy}.
\newblock Redactable blockchain using enhanced chameleon hash function.
\newblock In {\em 2019 5th International Conference on Advanced Computing
  Communication Systems (ICACCS)}, pages 323--328, 2019.

\bibitem{ateniese2017redactable}
G.~Ateniese, B.~Magri, D.~Venturi, and E.~Andrade.
\newblock Redactable blockchain--or--rewriting history in bitcoin and friends.
\newblock In {\em 2017 IEEE European Symposium on Security and Privacy
  (EuroS\&P)}, pages 111--126. IEEE, 2017.

\bibitem{buchman2016tendermint}
E.~Buchman.
\newblock {\em Tendermint: Byzantine fault tolerance in the age of
  blockchains}.
\newblock PhD thesis, 2016.

\bibitem{buterin2014next}
V.~Buterin et~al.
\newblock A next-generation smart contract and decentralized application
  platform.
\newblock {\em white paper}, 3(37), 2014.

\bibitem{Chandra}
T.~D. Chandra and S.~Toueg.
\newblock Unreliable failure detectors for reliable distributed systems.
\newblock {\em J. ACM}, 43(2):225–267, Mar. 1996.

\bibitem{deuber2019redactable}
D.~Deuber, B.~Magri, and S.~A.~K. Thyagarajan.
\newblock Redactable blockchain in the permissionless setting.
\newblock In {\em 2019 IEEE Symposium on Security and Privacy (SP)}, pages
  124--138. IEEE, 2019.

\bibitem{eyal2016bitcoin}
I.~Eyal, A.~E. Gencer, E.~G. Sirer, and R.~Van~Renesse.
\newblock Bitcoin-ng: A scalable blockchain protocol.
\newblock In {\em 13th $\{$USENIX$\}$ symposium on networked systems design and
  implementation ($\{$NSDI$\}$ 16)}, pages 45--59, 2016.

\bibitem{gilad2017algorand}
Y.~Gilad, R.~Hemo, S.~Micali, G.~Vlachos, and N.~Zeldovich.
\newblock Algorand: Scaling byzantine agreements for cryptocurrencies.
\newblock In {\em Proceedings of the 26th Symposium on Operating Systems
  Principles}, pages 51--68, 2017.

\bibitem{tezos}
L.~Goodman.
\newblock Tezos: A self-amending crypto-ledger position paper.
\newblock {\em Aug}, 3:2014, 2014.

\bibitem{FastFabric}
C.~{Gorenflo}, S.~{Lee}, L.~{Golab}, and S.~{Keshav}.
\newblock Fastfabric: Scaling hyperledger fabric to 20,000 transactions per
  second.
\newblock In {\em 2019 IEEE International Conference on Blockchain and
  Cryptocurrency (ICBC)}, pages 455--463, 2019.

\bibitem{zcash}
D.~Hopwood, S.~Bowe, T.~Hornby, and N.~Wilcox.
\newblock Zcash protocol specification.
\newblock {\em GitHub: San Francisco, CA, USA}, 2016.

\bibitem{couchDB}
https://couchdb.apache.org/.
\newblock Seamless multi-master sync, that scales from big data to mobile, with
  an intuitive http/json api and designed for reliability.

\bibitem{goLevelDB}
https://github.com/syndtr/goleveldb.
\newblock An implementation of the leveldb key/value database in the go
  programming language.

\bibitem{kogias2016enhancing}
E.~K. Kogias, P.~Jovanovic, N.~Gailly, I.~Khoffi, L.~Gasser, and B.~Ford.
\newblock Enhancing bitcoin security and performance with strong consistency
  via collective signing.
\newblock In {\em 25th $\{$usenix$\}$ security symposium ($\{$usenix$\}$
  security 16)}, pages 279--296, 2016.

\bibitem{Chameleon}
H.~Krawczyk and T.~Rabin.
\newblock Chameleon hashing and signatures, 1998.
\newblock Appeared in the THEORY OF CRYPTOGRAPHY LIBRARY and has been included
  in the ePrint Archive. talr@watson.ibm.com 10500 received March 17th, 1998.

\bibitem{8946222}
M.~{Kuzlu}, M.~{Pipattanasomporn}, L.~{Gurses}, and S.~{Rahman}.
\newblock Performance analysis of a hyperledger fabric blockchain framework:
  Throughput, latency and scalability.
\newblock In {\em 2019 IEEE International Conference on Blockchain
  (Blockchain)}, pages 536--540, 2019.

\bibitem{nakamotobitcoin}
S.~Nakamoto.
\newblock Bitcoin: A peer-to-peer electronic cash system.

\bibitem{8525390}
E.~{Palm}, O.~{Schelén}, and U.~{Bodin}.
\newblock Selective blockchain transaction pruning and state derivability.
\newblock In {\em 2018 Crypto Valley Conference on Blockchain Technology
  (CVCBT)}, pages 31--40, 2018.

\bibitem{parliament2016regulation}
E.~Parliament.
\newblock Regulation (eu) 2016/679 of the european parliament and of the coucil
  of 27 april 2016 on the protection of natural persons with regard to the
  processing of personal data and on the free movement of such data, and
  repealing directive 95/46/ec (general data protection regulation).
\newblock {\em Official Journal of the European Union L}, 119:1--88, 2016.

\bibitem{8526892}
P.~{Thakkar}, S.~{Nathan}, and B.~{Viswanathan}.
\newblock Performance benchmarking and optimizing hyperledger fabric blockchain
  platform.
\newblock In {\em 2018 IEEE 26th International Symposium on Modeling, Analysis,
  and Simulation of Computer and Telecommunication Systems (MASCOTS)}, pages
  264--276, 2018.

\bibitem{8876647}
N.~B. {Truong}, K.~{Sun}, G.~M. {Lee}, and Y.~{Guo}.
\newblock Gdpr-compliant personal data management: A blockchain-based solution.
\newblock {\em IEEE Transactions on Information Forensics and Security},
  15:1746--1761, 2020.

\bibitem{Visa}
Visa.
\newblock Visa fact sheet, 2018.

\bibitem{wirth2018privacy}
C.~Wirth and M.~Kolain.
\newblock Privacy by blockchain design: a blockchain-enabled gdpr-compliant
  approach for handling personal data.
\newblock In {\em Proceedings of 1st ERCIM Blockchain Workshop 2018}. European
  Society for Socially Embedded Technologies (EUSSET), 2018.

\bibitem{hotstuff}
M.~Yin, D.~Malkhi, M.~K. Reiter, G.~G. Gueta, and I.~Abraham.
\newblock Hotstuff: Bft consensus with linearity and responsiveness.
\newblock In {\em Proceedings of the 2019 ACM Symposium on Principles of
  Distributed Computing}, PODC '19, page 347–356, New York, NY, USA, 2019.
  Association for Computing Machinery.

\end{thebibliography}

\end{document}